\documentclass[conference]{IEEEtran}
\IEEEoverridecommandlockouts

\usepackage{cite}
\usepackage{amsmath,amssymb,amsfonts}
\usepackage{algorithmic}
\usepackage{graphicx}
\usepackage{textcomp}
\usepackage{xcolor}
\usepackage{comment}
\usepackage{multirow}
\usepackage{makecell}
\usepackage[normalem]{ulem}
\usepackage{tabularx}
\usepackage{booktabs}
\usepackage{amssymb}
\usepackage{url}
\def\BibTeX{{\rm B\kern-.05em{\sc i\kern-.025em b}\kern-.08em
    T\kern-.1667em\lower.7ex\hbox{E}\kern-.125emX}}
\begin{document}

\title{WaveRVQ: Iterative Residual Vector Quantization in the Waveform Domain for Neural Audio Coding\\
}

\author{
\IEEEauthorblockN{
Jin Sakuma$^{\star}$ \qquad
Takashi Maekaku$^{\star}$ \qquad
Keita Goto$^{\star}$ \\
Jinchuan Tian$^{\dagger}$ \qquad
Yusuke Shinohara$^{\star}$ \qquad
Shinji Watanabe$^{\dagger}$
}
\IEEEauthorblockA{
$^{\star}$ LY Corporation, Tokyo, Japan \qquad
$^{\dagger}$ Carnegie Mellon University, PA, USA
}
}

\maketitle

\begin{abstract}
Residual Vector Quantization (RVQ) is widely used in neural codecs to improve quantization accuracy through multi-stage residual quantization in the latent space. We propose WaveRVQ, which recursively reconstructs waveforms by computing residuals in the waveform domain rather than the latent space. At each iteration, the residual between the input and reconstructed waveforms is processed by the encoder, quantizer, and decoder, and the final waveform is obtained by summing the outputs from all iterations. By recursively reconstructing waveforms and directly optimizing the reconstruction at each iteration, WaveRVQ achieves more accurate reconstruction than conventional RVQ. Experiments show improvements in most evaluation metrics over conventional RVQ. For joint speech coding and enhancement under noisy conditions, WaveRVQ improves the mean opinion score from 3.23 to 3.51 compared with RVQ. Furthermore, it also enhances speech language model (SpeechLM)-based text-to-speech synthesis, reducing the word error rate from 15.1\% to 10.8\%.
\end{abstract}

\begin{IEEEkeywords}
Neural audio codec, residual vector quantization, speech coding, speech language model.
\end{IEEEkeywords}

\section{Introduction}

Conventional audio codecs~\cite{valin2012rfc, dietz2015overview} rely on sophisticated signal processing techniques and psychoacoustic knowledge.
In contrast, neural audio codecs~\cite{zeghidour2021soundstream,defossez2022high,kumar2023high,pmlr-v235-ju24b,zhang2024speechtokenizer,ji2025wavtokenizer,mousavi2025discrete,wu2024codec} enable end-to-end optimization and achieve high-quality compression at low bitrates. Recently, high-quality discrete representations produced by neural audio codecs have also been widely applied to speech language models and speech synthesis~\cite{borsos2023audiolm,tian2025espnet,wang2023neural,chen2024vall,du2024cosyvoice,arora2025on,cui-etal-2025-recent}. As a result, neural audio codecs have become a fundamental component of a wide range of speech and audio applications.

A neural audio codec typically consists of an encoder, a vector quantizer, and a decoder. The vector quantizer maps encoder outputs to the nearest entries in a learnable codebook. 
Recent neural audio codecs widely adopt residual vector quantization (RVQ)~\cite{gray1984vector,vasuki2006review}, which improves reconstruction quality at low bitrates by progressively quantizing latent residuals using multiple codebook stages.

In this work, we propose WaveRVQ, which iteratively computes residuals directly in the waveform domain (i.e., the time domain) rather than in the latent space. 
Motivated by the \textit{superposition principle} that a speech signal can be represented as the sum of multiple waveform components, the WaveRVQ framework reconstructs the target waveform by accumulating waveform components generated at successive iterations. 
Unlike typical RVQ-based codecs, which execute the encoding-quantization-decoding pipeline only once, WaveRVQ-based codecs repeatedly apply the same pipeline to progressively refine the waveform reconstruction.
Specifically, at each iteration, the residual waveform, defined as the difference between the input waveform and the reconstructed waveform, is processed by an encoder, quantizer, and decoder. 
Furthermore, because the reconstruction loss is computed on the reconstructed waveform obtained by summing the outputs from all iterations, each iteration is encouraged to generate a waveform component that directly contributes to the final reconstruction. This design improves the reconstruction quality at every iteration, enabling WaveRVQ to outperform conventional latent-space RVQ.

The main contributions of this study are as follows:
\begin{itemize}
\item We propose WaveRVQ, a framework for neural audio codecs that performs residual computation directly in the waveform domain. The model iteratively applies an encoder, quantizer, and decoder to progressively reconstruct waveforms.
\item Experimental results demonstrate that WaveRVQ consistently outperforms conventional RVQ in reconstruction quality. Furthermore, a detailed analysis reveals the contribution of each iteration to the reconstruction performance.
\end{itemize}

To facilitate reproducible research, we will release the source code upon acceptance.

\section{Related Work}
Neural audio codecs have evolved significantly since the early studies on neural speech coding~\cite{Morishima1990SpeechCB,kankanahalli2018end}. 
Recent neural audio codecs are typically based on the Vector Quantized Variational Autoencoder (VQ-VAE) framework~\cite{van2017neural}. VQ-VAE introduces a vector quantization framework with a learnable codebook and consists of an encoder, a vector quantizer, and a decoder.
Representative examples include SoundStream~\cite{zeghidour2021soundstream}, EnCodec~\cite{defossez2022high}, and the Descript Audio Codec (DAC)~\cite{kumar2023high}, all of which adopt RVQ as their quantization strategy. RVQ improves quantization performance through its multi-stage residual structure. Several extensions of RVQ have subsequently been proposed~\cite{gu2024esc,yang2023hifi,siuzdak2024snac}. Nevertheless, most existing RVQ-based neural audio codecs perform multi-stage residual quantization in the latent space.

Residual computation in the waveform domain has been explored in earlier hybrid speech codecs~\cite{zhen2019cascaded,zhen2020efficient,zhen2021scalable}. 
These studies primarily focused on speech communication and employed linear predictive coding (LPC) to improve computational efficiency.
However, residual computation in the waveform domain has been less explored in the context of modern neural audio codecs, such as SoundStream, EnCodec, and DAC, which are widely used in speech synthesis and speech enhancement.
In this work, we introduce waveform-domain residual computation into a fully end-to-end neural audio codec. The proposed method is applicable not only to speech coding but also to joint speech coding and enhancement, as well as speech language model (SpeechLM)-based text-to-speech synthesis, substantially extending the applicability of waveform-domain residual computation beyond prior work.

\section{WaveRVQ}
\label{section:method}
\subsection{Conventional RVQ in the Latent Space}
Let the original input waveform be denoted by 
$\mathbf{x} \in \mathbb{R}^{T}$, where $T$ denotes the waveform length. 
In conventional RVQ-based codecs, an encoder $\mathrm{Enc}(\cdot)$ maps $\mathbf{x}$ to a latent representation $\mathbf{Z}$, and a decoder $\mathrm{Dec}(\cdot)$ reconstructs the waveform from the quantized representation:
\begin{align}
  \mathbf{Z} &= \mathrm{Enc}(\mathbf{x}) \in \mathbb{R}^{d \times T_{\mathrm{f}}}, \label{eq:rvq_encode} \\
  \hat{\mathbf{x}} &= \mathrm{Dec}(\mathrm{RVQ}(\mathbf{Z})) \in \mathbb{R}^{T}, \label{eq:rvq_decode}
\end{align}
where $d$ denotes the feature dimension and $T_{\mathrm{f}}$ denotes the latent sequence length.
RVQ iteratively performs residual computation in the latent space using $n_{\mathrm{code}}$ sequential vector quantizers:
\begin{align}
  \mathrm{RVQ}(\mathbf{Z}) &= \sum_{m=1}^{n_{\mathrm{code}}} \hat{\mathbf{Z}}_{m}, \quad
  \hat{\mathbf{Z}}_{m} = \mathrm{VQ}_{m}(\mathbf{R}_{m}),
  \label{eq:rvq} \\
  \mathbf{R}_{m} &=
  \begin{cases}
  \mathbf{Z}, & m = 1, \\
  \mathbf{Z} - \displaystyle\sum_{j=1}^{m-1} 
  \hat{\mathbf{Z}}_{j}, & m > 1,
  \end{cases}  
  \label{eq:residual_subtraction}
\end{align}
where $\hat{\mathbf{Z}}_{m} \in \mathbb{R}^{d \times T_{\mathrm{f}}}$, $\mathbf{R}_{m} \in \mathbb{R}^{d \times T_{\mathrm{f}}}$, $n_{\mathrm{code}}$, and $\mathrm{VQ}_{m}(\cdot)$ denote the quantized output, the residual, number of codebooks, and the vector quantizer at the $m$-th stage, respectively.

\begin{figure}[t]
  \centering
  \includegraphics[width=\linewidth]{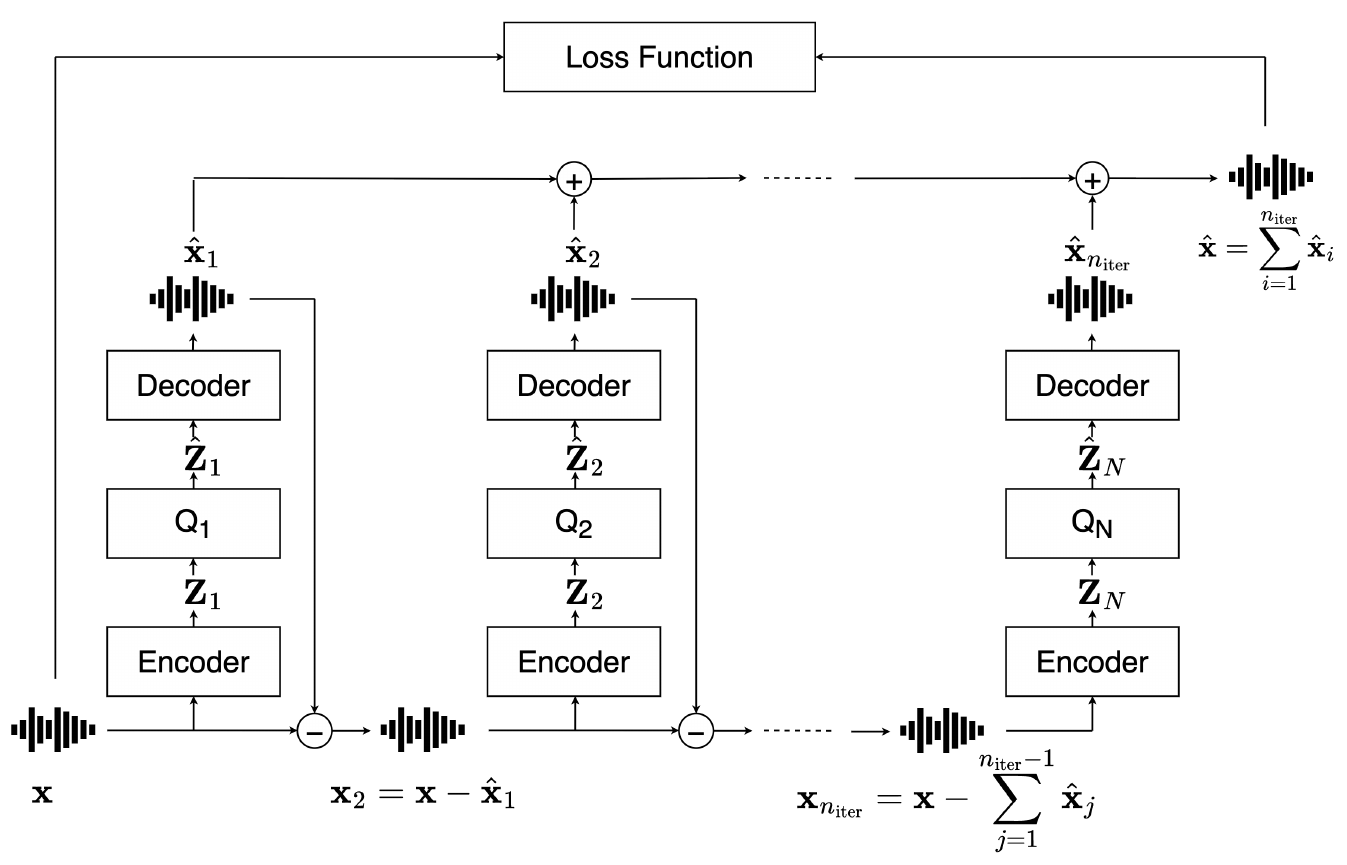}
  \caption{Overview of WaveRVQ. The encoder–quantizer–decoder stack is iteratively applied to waveform residuals in the waveform domain. The encoder and decoder share their parameters across all iterations. The final output is obtained by summing the reconstructed waveforms from all iterations.}
  \label{fig:architecture}
\end{figure}

\begin{table*}[t]
    \centering
    \caption{Comparison of joint speech coding and enhancement performance for different numbers of WaveRVQ iterations, together with the RVQ-based SoundStream at 6 kbps on the LRAC Challenge 2025 dataset. The best scores are shown in \textbf{bold}, and scores outperforming RVQ-based model are highlighted in \colorbox{blue!10}{blue}.}
    \scalebox{0.71}{
    \begin{tabular}{l|c|cc|ccccccccc}
    \toprule
        Condition & VQ Method & $n_{\mathrm{iter}}$ & $n_{\mathrm{code}}$ & PESQ ($\uparrow$) & UTMOS ($\uparrow$) & Audiobox AE-CE ($\uparrow$) & ScoreQ-ref ($\downarrow$) & Sheet-SSQA ($\uparrow$) & WER [\%] ($\downarrow$) & SPK-Sim ($\uparrow$) \\
        \midrule        
        \multirow{4}{*}{Clean} & RVQ & 1 & 6 & 2.17 &3.27 &5.44 &0.353 &3.81 &3.4	&0.704   \\
        \cmidrule(lr){2-11}
                              & \multirow{3}{*}{WaveRVQ} & 2 & 3 & \colorbox{blue!10}{2.18} &3.24 &5.38 &0.364 &\colorbox{blue!10}{3.81} &3.7 &0.697 \\                              
                              && 3 & 2 & \colorbox{blue!10}{\textbf{2.48}} &\colorbox{blue!10}{\textbf{3.33}} &\colorbox{blue!10}{\textbf{5.46}} &\colorbox{blue!10}{\textbf{0.317}} &\colorbox{blue!10}{\textbf{3.97}} &3.8 &\colorbox{blue!10}{0.785}    \\
                              & & 6 & 1 & \colorbox{blue!10}{2.32} & 3.12 & 5.33 & 0.398 & 3.80 & \colorbox{blue!10}{\textbf{3.3}} & \colorbox{blue!10}{\textbf{0.813}} \\
        \hline
        \multirow{4}{*}{Noisy} & RVQ & 1 & 6 & 1.74 & 2.74 & 4.71 & 0.718 & 3.07 & 15.4 & 0.550 \\
        \cmidrule(lr){2-11}
                              & \multirow{3}{*}{WaveRVQ} & 2 & 3 & \colorbox{blue!10}{1.76} & \colorbox{blue!10}{2.77} & \colorbox{blue!10}{\textbf{4.73}} & \colorbox{blue!10}{0.699} & \colorbox{blue!10}{3.14} & 16.5 & \colorbox{blue!10}{0.552} \\
                              & & 3 & 2 & \colorbox{blue!10}{\textbf{1.88}} & \colorbox{blue!10}{\textbf{2.81}} & \colorbox{blue!10}{4.72} & \colorbox{blue!10}{\textbf{0.691}} & \colorbox{blue!10}{\textbf{3.20}} & \colorbox{blue!10}{12.7} & \colorbox{blue!10}{0.636} \\
                              & & 6 & 1 & \colorbox{blue!10}{1.79} & 2.59 & 4.56 & 0.775 & 3.03 & \colorbox{blue!10}{\textbf{12.5}} & \colorbox{blue!10}{\textbf{0.652}} \\
        \hline
        \multirow{4}{*}{Reverb} & RVQ & 1 & 6 & 1.27 & 1.55 & 4.12 & 0.900 & 2.79 & 21.0 & 0.547 \\
        \cmidrule(lr){2-11}
                              & \multirow{3}{*}{WaveRVQ}& 2 & 3 & \colorbox{blue!10}{1.27} & \colorbox{blue!10}{\textbf{1.60}} & \colorbox{blue!10}{4.18} & 0.919 & \colorbox{blue!10}{2.81} & 23.0 & \colorbox{blue!10}{0.556} \\
                              & & 3 & 2 & \colorbox{blue!10}{\textbf{1.30}} & 1.53 & \colorbox{blue!10}{4.23} & \colorbox{blue!10}{0.868} & \colorbox{blue!10}{3.10} & \colorbox{blue!10}{15.4} & \colorbox{blue!10}{0.658} \\
                              & & 6 & 1 & \colorbox{blue!10}{1.28} & 1.49 & \colorbox{blue!10}{\textbf{4.29}} & \colorbox{blue!10}{\textbf{0.825}} & \colorbox{blue!10}{\textbf{3.17}} & \colorbox{blue!10}{\textbf{13.1}} & \colorbox{blue!10}{\textbf{0.702}} \\
        \bottomrule
    \end{tabular}}
    \label{tab:iterations}
\end{table*}

\subsection{RVQ in the Waveform Domain}
\label{section:wavervq}
Fig.~\ref{fig:architecture} illustrates the architecture of WaveRVQ. In the WaveRVQ framework, an encoder, a vector quantizer, and a decoder are applied iteratively.
At each iteration, similarly to conventional RVQ-based codecs described in Eqs.~\eqref{eq:rvq_encode} and~\eqref{eq:rvq_decode}, the encoder and decoder process the input residual waveform.
Specifically, in the $i$-th iteration, the input waveform $\mathbf{x}_{i} \in \mathbb{R}^{T}$ is encoded into a latent representation $\mathbf{Z}_{i}$ by the encoder $\mathrm{Enc}(\cdot)$.
The decoder $\mathrm{Dec}(\cdot)$ then reconstructs the waveform $\hat{\mathbf{x}}_{i}$ from a quantized latent representation produced by a generic quantizer $\mathrm{Q}_{i}(\cdot)$:
\begin{align}
  \mathbf{Z}_{i} &= \mathrm{Enc}(\mathbf{x}_{i}) \in \mathbb{R}^{d \times T_{\mathrm{f}}}, \label{eq:encode} \\
  \hat{\mathbf{x}}_{i} &= \mathrm{Dec}(\mathrm{Q}_{i}(\mathbf{Z}_{i})) \in \mathbb{R}^{T}, \label{eq:decode}
\end{align}
where $\mathrm{Q}_{i}$ denotes a generic vector quantizer with $n_{\mathrm{code}}$ codebooks.
In conventional RVQ, the final quantized latent representation is obtained by summing the outputs of all vector quantizers, as shown in Eq.~\eqref{eq:rvq}. In contrast, WaveRVQ performs the summation in the waveform domain.
The final reconstructed waveform $\hat{\mathbf{x}}$ is obtained by summing the reconstructed waveforms from all $n_{\mathrm{iter}}$ iterations:
\begin{align}
  \hat{\mathbf{x}} = \sum_{i=1}^{n_{\mathrm{iter}}} \hat{\mathbf{x}}_{i} \in \mathbb{R}^{T}.
  \label{eq:final_output}
\end{align}
While conventional RVQ defines the residual at the $m$-th stage in the latent space as in Eq.~\eqref{eq:residual_subtraction}, WaveRVQ defines the residual waveform $\mathbf{x}_{i}$, which serves as the input to the $i$-th iteration, as
\begin{align}
  \mathbf{x}_{i} =
  \begin{cases}
    \mathbf{x}, & i = 1, \\
    \mathbf{x} - \displaystyle\sum_{j=1}^{i-1} \hat{\mathbf{x}}_{j}, & i > 1,
  \end{cases}
  \label{eq:residual}
\end{align}
where $\hat{\mathbf{x}}_{j} \in \mathbb{R}^{T}$ denotes the reconstructed waveform at iteration $j$.

The loss function described in Section~\ref{section:training_objective} is computed on $\hat{\mathbf{x}}$. As implied by Eqs.~\eqref{eq:final_output} and~\eqref{eq:residual}, the output waveform at each iteration is regularized, which helps improve training stability.

The WaveRVQ framework applies the encoder and decoder for $n_{\mathrm{iter}}$ iterations. The encoder and decoder share their parameters across all iterations.
As a result, for the same number of codebooks, increasing the number of iterations does not increase the model size compared with the baseline.
In contrast, the computational cost increases approximately linearly with the number of iterations $n_{\mathrm{iter}}$.
However, this increase in computational cost is not necessarily problematic for some downstream tasks.
As shown in Section~\ref{section:speechlm}, for the SpeechLM-based text-to-speech (TTS) task, the additional processing time introduced by the proposed method does not become a bottleneck in the overall system.

\subsection{Training Objectives}
\label{section:training_objective}
WaveRVQ is optimized within a generative adversarial network (GAN) framework. Specifically, we employ the least-squares GAN (LSGAN)~\cite{mao2017least} objective for adversarial training.
The discriminator $\mathcal{D}$ is trained to distinguish a real waveform $\mathbf{x}$ from a generated waveform $\hat{\mathbf{x}}$ as:
\begin{align}
    \mathcal{L}_{\mathcal{D}} = \mathbb{E}_{\mathbf{x}} \left[(\mathcal{D}(\mathbf{x}) - 1)^2 \right] + \mathbb{E}_{\hat{\mathbf{x}}} \left[\mathcal{D}(\hat{\mathbf{x}})^2 \right].
    \label{eq:d_loss}
\end{align}
Following EnCodec~\cite{defossez2022high}, we employ a multi-scale short-time Fourier transform (MS-STFT) discriminator. 
The generator adversarial loss is defined as:
\begin{align}
    \mathcal{L}_{\mathrm{adv}} = \mathbb{E}_{\hat{\mathbf{x}}} \left[(\mathcal{D}(\hat{\mathbf{x}}) - 1)^2 \right].
    \label{eq:adv_loss}
\end{align}
In addition, the overall objective of the generator $\mathcal{G}$ includes the following losses: commitment loss~\cite{van2017neural}, waveform L1 reconstruction loss, mel-spectrogram loss~\cite{yamamoto2020parallel}, and discriminator feature matching loss. The generator loss $\mathcal{L}_{\mathcal{G}}$ is defined as the weighted sum of these loss terms.

The codebook is updated using an exponential moving average. In addition, gradients are backpropagated through the codebook lookup using the straight-through estimator~\cite{bengio2013estimating}. 
WaveRVQ framework reconstructs a waveform component at each iteration, and the commitment loss is averaged across all iterations. The other loss functions are applied to the final output waveform $\hat{\mathbf{x}}$ defined in Eq.~\eqref{eq:final_output}.
The overall training objective can be formulated as the following min–max optimization:
\begin{align}    
    \min_{\mathcal{G}} \max_{\mathcal{D}} \ \mathcal{L}(\mathcal{G},\mathcal{D}) = \mathcal{L}_{\mathcal{G}} - \lambda_{\mathrm{adv}} \cdot \mathcal{L}_{\mathcal{D}},
\end{align}
where $\lambda_{\mathrm{adv}}$ denotes the weighting parameter for the adversarial loss.

As shown in Eq.~\eqref{eq:final_output}, the final output $\hat{\mathbf{x}}$ is computed as the sum of the outputs from all iterations, $\hat{\mathbf{x}}_{i}$. Because the loss functions in Eq.~\eqref{eq:adv_loss} and the other generator loss functions are applied to this final output, they regularize the waveform generated at each iteration.
This enables waveform-domain optimization at every iteration, which is a key advantage of WaveRVQ and contributes to improved reconstruction performance.

\begin{table*}[t]
    \centering
    \caption{Speech coding performance of WaveRVQ and the RVQ-based SoundStream, EnCodec, and DAC on LRAC Challenge 2025 dataset. The scores outperforming the RVQ-based model are highlighted in \colorbox{blue!10}{blue}. }
    \scalebox{0.71}{
    \begin{tabular}{l|c|c|cc|ccccccccc}
    \toprule
        Codec & Condition & VQ Method & $n_{\mathrm{iter}}$ & $n_{\mathrm{code}}$ & PESQ ($\uparrow$) & UTMOS ($\uparrow$) & Audiobox AE-CE ($\uparrow$) & ScoreQ-ref ($\downarrow$) & Sheet-SSQA ($\uparrow$) & WER [\%] ($\downarrow$) & SPK-Sim ($\uparrow$) \\
        \midrule        
        \multirow{6}{*}{SoundStream} & \multirow{2}{*}{Clean} & RVQ & 1 & 6 & 2.82 & 3.15 & 5.31 & 0.391 & 3.75 & 2.2 & 0.842 \\   
                             & & WaveRVQ & 2 & 3 & \colorbox{blue!10}{2.87} & \colorbox{blue!10}{3.28} & \colorbox{blue!10}{5.36} & \colorbox{blue!10}{0.371} & \colorbox{blue!10}{3.81} & 2.4 & \colorbox{blue!10}{0.852} \\
        \cmidrule(lr){2-12}
        & \multirow{2}{*}{Noisy} & RVQ & 1 & 6 & 2.67 & 1.80 & 4.04 & 0.299 & 2.48 & 7.3 & 0.800 \\        
                               & & WaveRVQ & 2 & 3 & 2.60 & \colorbox{blue!10}{1.84} & \colorbox{blue!10}{4.11} & 0.305 & \colorbox{blue!10}{2.49} & 7.5 & \colorbox{blue!10}{0.814} \\
        \cmidrule(lr){2-12}
        & \multirow{2}{*}{Reverb} & RVQ & 1 & 6 & 2.88 & 1.32 & 4.17 & 0.298 & 2.99 & 7.0 & 0.844 \\        
                                & & WaveRVQ & 2 & 3 & 2.76 & \colorbox{blue!10}{1.33} & \colorbox{blue!10}{4.30} & \colorbox{blue!10}{0.255} & \colorbox{blue!10}{3.03} & 7.9 & \colorbox{blue!10}{0.861}\\
        
        \hline
        
        \multirow{6}{*}{EnCodec} & \multirow{2}{*}{Clean} & RVQ & 1 & 6 & 2.07 & 2.77 & 5.17 & 0.503 & 3.33 & 2.8 & 0.692\\   
                                   & & WaveRVQ & 2 & 3 & \colorbox{blue!10}{2.53} & \colorbox{blue!10}{3.40} & \colorbox{blue!10}{5.55} & \colorbox{blue!10}{0.277} & \colorbox{blue!10}{3.95} & 3.7 & \colorbox{blue!10}{0.821} \\
        \cmidrule(lr){2-12}
        & \multirow{2}{*}{Noisy} & RVQ & 1 & 6 & 1.69 & 1.66 & 3.91 & 0.409 & 2.31 & 11.9 & 0.631\\        
                                 & & WaveRVQ & 2 & 3 & \colorbox{blue!10}{2.01} & \colorbox{blue!10}{1.94} & \colorbox{blue!10}{4.18} & \colorbox{blue!10}{0.282} & \colorbox{blue!10}{2.54} & \colorbox{blue!10}{8.4} & \colorbox{blue!10}{0.762}\\
        \cmidrule(lr){2-12}
        & \multirow{2}{*}{Reverb} & RVQ & 1 & 6 & 1.56 & 1.34 & 4.09 & 0.366 & 2.79 & 15.0 & 0.669 \\        
                                  & & WaveRVQ & 2 & 3 & \colorbox{blue!10}{1.68} & \colorbox{blue!10}{1.40} & \colorbox{blue!10}{4.30} & \colorbox{blue!10}{0.244} & \colorbox{blue!10}{3.05} & \colorbox{blue!10}{8.8} & \colorbox{blue!10}{0.807} \\
        
        \hline
        
        \multirow{6}{*}{DAC} & \multirow{2}{*}{Clean} & RVQ & 1 & 6 & 3.22 & 3.68 & 5.57 & 0.217 & 4.10 & 2.4 & 0.791 \\   
                             & & WaveRVQ & 3 & 2 & 3.05 & \colorbox{blue!10}{3.76} & \colorbox{blue!10}{5.62} & \colorbox{blue!10}{0.189} & \colorbox{blue!10}{4.18} & 2.6 & \colorbox{blue!10}{0.826} \\
        \cmidrule(lr){2-12}
        & \multirow{2}{*}{Noisy} & RVQ & 1 & 6 & 2.70 & 2.02 & 4.20 & 0.250 & 2.63 & 8.4 & 0.756 \\        
                               & & WaveRVQ & 3 & 2 & 2.56 & \colorbox{blue!10}{2.10} & \colorbox{blue!10}{4.29} & \colorbox{blue!10}{0.243} & \colorbox{blue!10}{2.67} & 8.6 & \colorbox{blue!10}{0.781} \\
        \cmidrule(lr){2-12}
        & \multirow{2}{*}{Reverb} & RVQ & 1 & 6 & 2.61 & 1.35 & 4.25 & 0.275 & 2.99 & 7.7 & 0.773\\        
                                & & WaveRVQ & 3 & 2 & 2.35 & \colorbox{blue!10}{1.39} & \colorbox{blue!10}{4.27} & \colorbox{blue!10}{0.270} & \colorbox{blue!10}{3.04} & 9.8 & \colorbox{blue!10}{0.801}\\
        \bottomrule
    \end{tabular}}
    \label{tab:other_architecture}
\end{table*}

\begin{table*}[t]
    \centering
    \caption{Joint speech coding and enhancement performance of WaveRVQ and baseline RVQ-based EnCodec and DAC models on LRAC Challenge 2025 dataset. The scores outperforming the RVQ-based model are highlighted in \colorbox{blue!10}{blue}. }
    \scalebox{0.71}{
    \begin{tabular}{l|c|c|cc|ccccccccc}
    \toprule
        Codec & Condition & VQ Method & $n_{\mathrm{iter}}$ & $n_{\mathrm{code}}$ & PESQ ($\uparrow$) & UTMOS ($\uparrow$) & Audiobox AE-CE ($\uparrow$) & ScoreQ-ref ($\downarrow$) & Sheet-SSQA ($\uparrow$) & WER [\%]($\downarrow$) & SPK-Sim ($\uparrow$)\\
        \midrule        
        \multirow{6}{*}{EnCodec} & \multirow{2}{*}{Clean} & RVQ & 1 & 6 & 2.15 & 3.16 & 5.50 & 0.339 & 3.75 & 3.5 & 0.673   \\   
                                   & & WaveRVQ & 3 & 2 & \colorbox{blue!10}{2.34} & \colorbox{blue!10}{3.29} & \colorbox{blue!10}{5.51} & \colorbox{blue!10}{0.319} & \colorbox{blue!10}{3.76} & 4.1 & \colorbox{blue!10}{0.746} \\
        \cmidrule(lr){2-12}
        & \multirow{2}{*}{Noisy} & RVQ & 1 & 6 & 1.78 & 2.82 & 4.96 & 0.595 & 3.23 & 14.9 & 0.539 \\        
                                 & & WaveRVQ & 3 & 2 & \colorbox{blue!10}{1.93} & \colorbox{blue!10}{2.93} & \colorbox{blue!10}{5.01} & \colorbox{blue!10}{0.569} & 3.20 & \colorbox{blue!10}{13.0} & \colorbox{blue!10}{0.607}\\
        \cmidrule(lr){2-12}
        & \multirow{2}{*}{Reverb} & RVQ & 1 & 6 & 1.31 & 1.85 & 4.61 & 0.709 & 3.14 & 17.3 & 0.564\\        
                                  & & WaveRVQ & 3 & 2 & \colorbox{blue!10}{1.36} & \colorbox{blue!10}{1.88} & \colorbox{blue!10}{4.62} & 0.760 & 3.05 & \colorbox{blue!10}{16.1} & \colorbox{blue!10}{0.624}\\
        \hline
        \multirow{6}{*}{DAC} & \multirow{2}{*}{Clean} & RVQ & 1 & 6 & 2.80 & 3.83 & 5.67 & 0.198 & 4.17 & 3.3 & 0.717   \\   
                             & & WaveRVQ & 2 & 3 & \colorbox{blue!10}{2.85} & \colorbox{blue!10}{3.85} & \colorbox{blue!10}{5.70} & \colorbox{blue!10}{0.186} & \colorbox{blue!10}{4.21} & \colorbox{blue!10}{3.0} & \colorbox{blue!10}{0.772}\\
        \cmidrule(lr){2-12}
        & \multirow{2}{*}{Noisy} & RVQ & 1 & 6 & 2.18 & 3.48 & 5.27 & 0.435 & 3.70 & \colorbox{blue!10}{11.5} & 0.603 \\        
                               & & WaveRVQ & 2 & 3 & \colorbox{blue!10}{2.22} & \colorbox{blue!10}{3.54} & \colorbox{blue!10}{5.35} & \colorbox{blue!10}{0.403} & \colorbox{blue!10}{3.78} & 11.8 & \colorbox{blue!10}{0.658} \\
        \cmidrule(lr){2-12}
        & \multirow{2}{*}{Reverb} & RVQ & 1 & 6 & 1.33 & 1.66 & 4.53 & 0.813 & 3.18 & 12.7 & 0.580 \\        
                                & & WaveRVQ & 2 & 3 & \colorbox{blue!10}{1.36} & \colorbox{blue!10}{1.75} & \colorbox{blue!10}{4.55} & \colorbox{blue!10}{0.773} & \colorbox{blue!10}{3.21} & \colorbox{blue!10}{12.4} & \colorbox{blue!10}{0.637} \\        
        \bottomrule
    \end{tabular}}
    \label{tab:other_architecture_joint}
\end{table*}

\section{Experiments}
We evaluate the proposed method on two tasks.

\textbf{Speech Coding}: The objective is to reconstruct the original speech waveform from its compressed representation. 

\textbf{Joint Speech Coding and Enhancement}: The objective is to reconstruct clean speech from noisy input while simultaneously performing speech compression. 

\subsection{Dataset}
We evaluate WaveRVQ on the publicly available Low-Resource Audio Codec (LRAC) Challenge 2025 dataset~\cite{wojcicki2025low}, enabling full experimental reproducibility. The training set consists of approximately 702.7 hours of speech data and 335 hours of noise data.
Following the LRAC Challenge 2025 setup, noise augmentation is applied with an 80\% probability, where the signal-to-noise ratio (SNR) is sampled uniformly from $-5$ to 30 dB. Reverberation augmentation is applied with a 50\% probability.
The evaluation set consists of three conditions: clean, noisy, and reverberant.

\subsection{Evaluation}
Evaluation is conducted using the Versa toolkit~\cite{shi2025versa}. Following the LRAC Challenge 2025, we employ the following evaluation metrics: the non-matching reference mode of Speech Contrastive Regression for Quality (ScoreQ-ref)~\cite{ragano2024scoreq}, UTMOS from the VoiceMOS Challenge 2022~\cite{saeki2022utmos}, Speech Human Evaluation Estimation Toolkit for Subjective Speech Quality Assessment (Sheet-SSQA)~\cite{huang2024mos}, Perceptual Evaluation of Speech Quality (PESQ)~\cite{rix2001perceptual}, and Audiobox Aesthetics Cross-Entropy (Audiobox AE-CE)~\cite{tjandra2025meta}. 
In addition, we evaluate intelligibility using word error rate (WER) computed with Whisper-large-v3~\cite{radford2023robust} and speaker identity preservation using speaker similarity (SPK-Sim)\footnote{We use the RawNet3-based speaker embedding model~\cite{jung2022pushing, jung2024espnet} available at \url{https://huggingface.co/espnet/voxcelebs12_rawnet3}.}. 
Since the LRAC Challenge 2025 dataset does not provide reference transcriptions for the test set, we generate transcriptions of the reference audio using Whisper-large-v3 and use them as the ground-truth transcripts for WER computation.

\subsection{Experimental Setting}
\label{subsection:setting}
We conduct our experiments using ESPNet-Codec~\cite{shi2024espnet}, the codec implementation of the ESPNet toolkit~\cite{watanabe2018espnet}.
As baseline models, we adopt three representative RVQ-based neural audio codecs: SoundStream~\cite{zeghidour2021soundstream}, EnCodec~\cite{defossez2022high}, and DAC~\cite{kumar2023high}. All models share the same training data and codebase, enabling a fair comparison. In addition, all implementations are open source, ensuring reproducibility.

For SoundStream, we use the same architecture as the Track 2 baseline model of the LRAC Challenge 2025~\cite{isik2025baseline}.
The encoder downsampling factor and hidden dimension $d$ are 240 and 512, respectively. The quantizer consists of an RVQ with $n_{\mathrm{code}}=6$, where each codebook contains 1024 entries. The quantizer operates on 24-dimensional features, which are projected to and from the quantization space using linear projection layers. The decoder mirrors the encoder architecture. The overall codec bitrate is 6.0 kbps.
For EnCodec and DAC, we use comparable configurations with an encoder downsampling factor of 320 and a hidden dimension $d$ of 512. Both models employ an RVQ with $n_{\mathrm{code}}=6$ and 1024 entries per codebook and operate at a bitrate of 4.5 kbps. The quantizer hidden dimension is set to 256 for EnCodec and 128 for DAC. In both cases, the decoder mirrors the encoder architecture.

We apply the WaveRVQ framework to the same encoder and decoder used in the corresponding baseline model. As described in Eq.~\eqref{eq:decode}, the vector quantizer in WaveRVQ is generic and can be implemented as either a single-codebook VQ or an RVQ. To ensure the same bitrate, the number of iterations $n_{\mathrm{iter}}$ and the number of codebooks $n_{\mathrm{code}}$ are set to $(n_{\mathrm{iter}}, n_{\mathrm{code}}) = \{(2, 3), (3, 2), (6, 1)\}$. When $n_{\mathrm{code}} > 1$, RVQ is used as the quantizer at each iteration as in the baseline models. 
Since all configurations satisfy $n_{\mathrm{iter}} \times n_{\mathrm{code}} = 6$, they use the same total number of codebooks and bitrate.

As described in Section~\ref{section:wavervq}, the proposed method shares the encoder and decoder parameters across iterations.
Hence, the WaveRVQ-based models have the same number of parameters as their corresponding baseline models: 1.8M, 22.5M, and 63.4M parameters for the SoundStream-, EnCodec-, and DAC-based models, respectively.

As defined in Sec.~\ref{section:training_objective}, the training objective consists of the commitment loss, waveform L1 reconstruction loss, mel-spectrogram loss, adversarial loss, and feature matching loss, weighted by 1.0, 1.0, 45.0, 1.0, and 2.0, respectively.
Training is conducted using four NVIDIA A100 GPUs with a batch size of 32. The SoundStream-based models are trained for 500 epochs, whereas the EnCodec- and DAC-based models are trained for 200 epochs.

\section{Results and Discussions}
\subsection{Impact of the Number of Iterations}
\label{section:results_iterations}
We incorporate WaveRVQ into the SoundStream architecture and compare its performance across different numbers of iterations $n_{\mathrm{iter}}$ for the joint speech coding and enhancement task.
As shown in Table~\ref{tab:iterations}, when $n_{\mathrm{iter}}=3$, WaveRVQ-based model outperforms the RVQ-based model on most evaluation metrics across all test sets. In particular, for the noisy and reverberant test sets, WaveRVQ-based model also outperforms the RVQ-based model on many evaluation metrics when $n_{\mathrm{iter}}=2$ and $6$.
Furthermore, WER and speaker similarity consistently improve as the number of iterations increases.
These results indicate that our method outperforms conventional latent-space RVQ by iteratively applying the encoder-quantizer-decoder pipeline and optimizing in the waveform domain at each iteration, as described in Section~\ref{section:training_objective}.

\begin{table}[t]
\centering
\caption{Subjective MOS under noisy and reverberant conditions. Values are reported as Mean $\pm$ 95\% Confidence Interval.}
\label{tab:mos_results}
\begin{tabular}{lcc}
\toprule
VQ Method & Noisy Speech & Reverberant Speech \\
\midrule
RVQ & 3.23 $\pm$ 0.12 & 2.13 $\pm$ 0.08 \\
WaveRVQ & \textbf{3.51} $\pm$ 0.11 & \textbf{2.32} $\pm$ 0.08 \\              
\bottomrule
\end{tabular}
\end{table}


\begin{figure}[t]
  \centering
  \includegraphics[width=\linewidth]{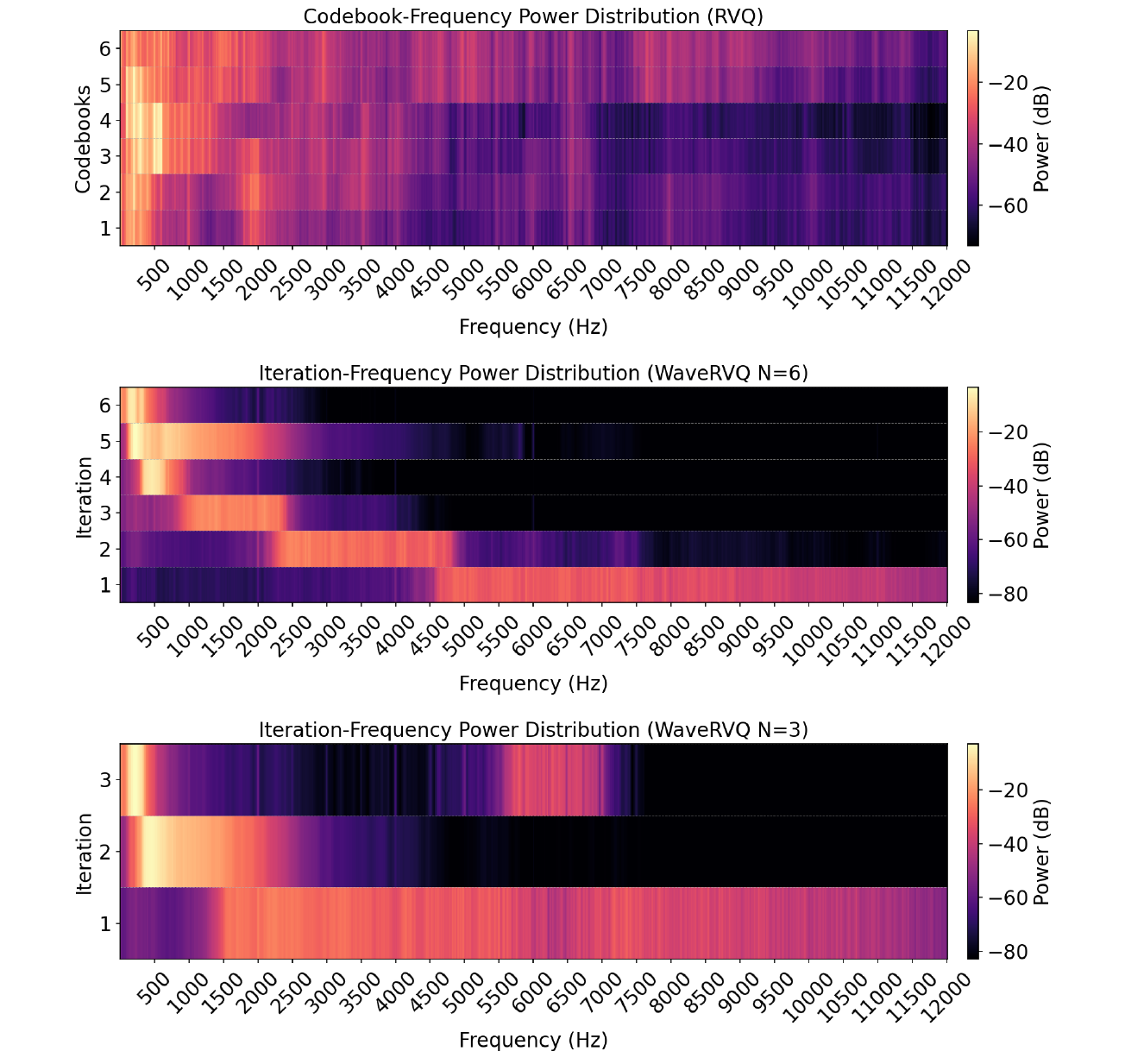}
  \caption{Frequency attention heatmaps of SoundStream using RVQ and WaveRVQ (($n_{\mathrm{iter}}, n_{\mathrm{code}}) = (6,1), (3,2)$) on the noisy test set.}
  \label{fig:analysis}
\end{figure}

\subsection{Evaluation on Multiple Architectures}
\label{section:results_multi_architecture}
To evaluate the generalizability of the proposed method, we apply WaveRVQ to SoundStream, EnCodec, and DAC, and compare it with conventional RVQ-based methods.
We select the WaveRVQ configuration $(n_{\mathrm{iter}}, n_{\mathrm{code}})$ from the candidate set $\{(2,3), (3,2), (6,1)\}$ based on its performance and adopt the best-performing configuration.

As summarized in Table~\ref{tab:other_architecture}, the WaveRVQ-based model outperforms the RVQ-based model on most evaluation metrics across all architectures in the speech coding task. In particular, for EnCodec, the WaveRVQ-based model significantly outperforms the RVQ-based model on all metrics for all test sets.

The same trend is observed in the joint speech coding and enhancement task, as shown in Tables~\ref{tab:iterations} and~\ref{tab:other_architecture_joint}. Across all architectures, the WaveRVQ-based model achieves better results than the RVQ-based method on most evaluation metrics.

These results confirm that the proposed method improves both speech coding and enhancement performance. Moreover, the consistent gains across multiple architectures suggest that the proposed method has strong generalizability.

\subsection{Subjective Evaluation}
We conduct an Absolute Category Rating (ACR)~\cite{ITU-T_P800_1996} evaluation to compare DAC models based on RVQ and WaveRVQ for the joint speech coding and enhancement task. 
WaveRVQ is configured with $n_{\mathrm{iter}} = 3$ and $n_{\mathrm{code}} = 2$. The evaluation data consist of 60 utterances, with 30 utterances randomly selected from each of the noisy and reverberant test sets of the LRAC Challenge 2025 Dataset. The listening test involves 15 non-expert participants who listen through headphones. 
As shown in Table~\ref{tab:mos_results}, the WaveRVQ-based model achieves a higher Mean Opinion Score (MOS) than the RVQ-based model on both test sets. Furthermore, paired $t$-tests yield $p$-values of 0.0004 and 0.007 for the noisy and reverberant test sets, respectively. These results confirm that the WaveRVQ-based model provides a statistically significant improvement over the RVQ-based model.

\begin{figure}[t]
  \centering  
  \includegraphics[width=\linewidth]{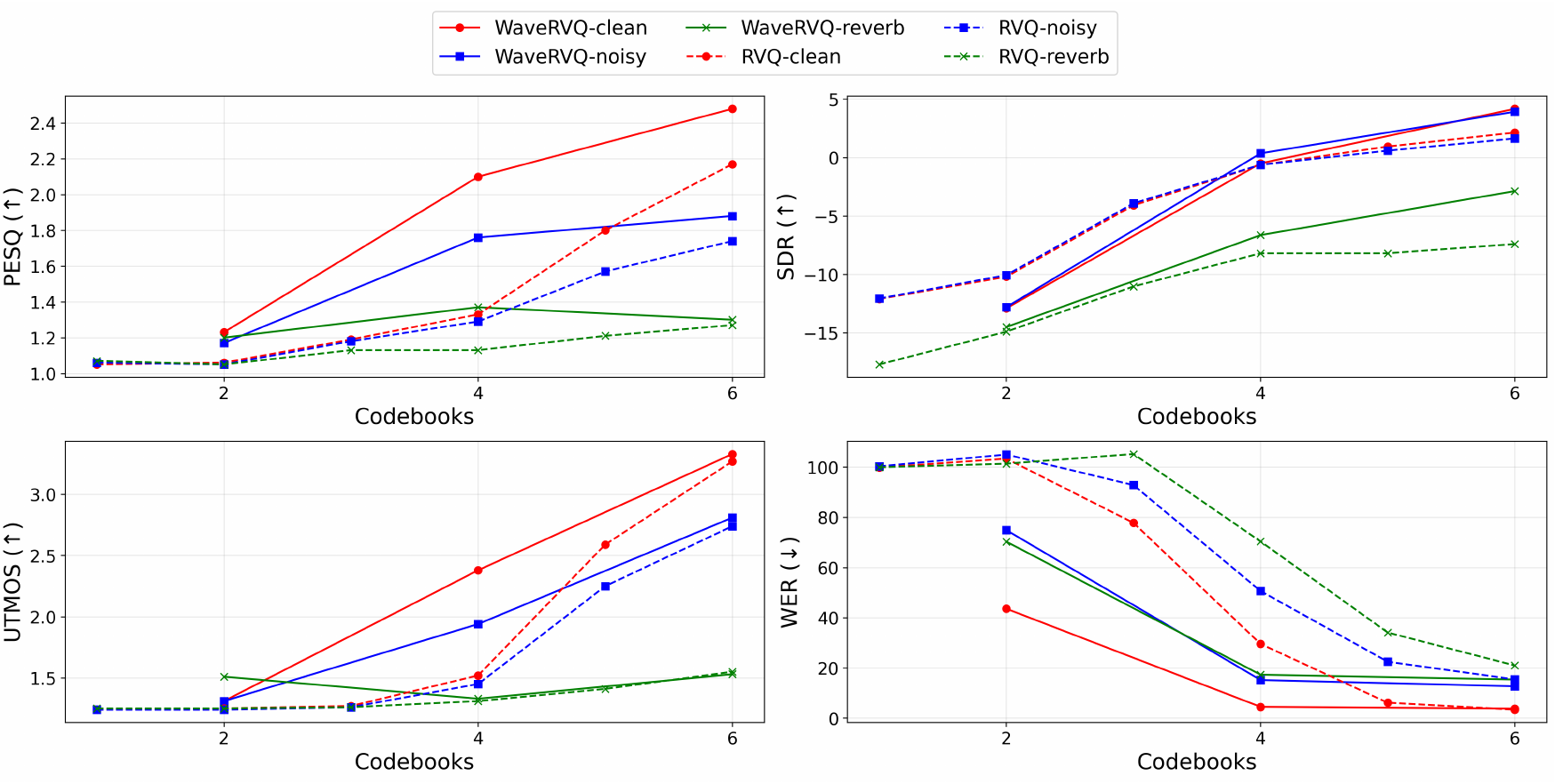}
  \caption{Evaluation metrics versus the number of codebooks for RVQ and WaveRVQ ($n_{\mathrm{iter}}=3$). Results are reported using the first $k$ codebooks ($k=1,\dots,6$) for joint speech coding and enhancement.}
  \label{fig:analysis_iteration_scores}
\end{figure}

\subsection{Analysis}
\label{section:analysis}
Fig.~\ref{fig:analysis} illustrates the frequency regions on which each RVQ codebook and each WaveRVQ iteration focuses in the joint speech coding and enhancement task. The backbone architecture is SoundStream.
For RVQ, the waveform corresponding to each codebook is reconstructed by feeding only the latent representation produced by the target codebook into the decoder. For WaveRVQ, the waveform generated at the target iteration is extracted.
For each test sample, we compute the magnitude spectrum in decibels, average it over the time axis, and then compute the median across all samples. The resulting values are visualized as a heatmap.
As similar trends are observed across all test sets, we present the results for the noisy test set.
Because RVQ operates on residuals in the latent feature space, the frequency-wise roles of the individual codebooks are not clearly identifiable.
In contrast, WaveRVQ exhibits clear frequency-wise specialization across iterations. Specifically, the range of 100–200 Hz, which corresponds to the typical pitch range of human speech~\cite{leung2022speaking}, is represented at a finer resolution, whereas higher frequencies are represented more coarsely.
This frequency allocation resembles the frequency resolution of human auditory perception. Since auditory frequency resolution is approximately constant on perceptual scales~\cite{zwicker1961subdivision,glasberg1990derivation}, effective bandwidth becomes narrower at lower frequencies and broader at higher frequencies.
Overall, these results suggest that WaveRVQ implicitly learns a perceptually aligned frequency decomposition.

Fig.~\ref{fig:analysis_iteration_scores} shows the WER, PESQ, UTMOS, and signal-to-distortion ratio (SDR) obtained by cumulatively using the first $k$ codebooks ($k=1,\ldots,6$). 
We added SDR to evaluate waveform reconstruction fidelity, in addition to the existing perceptual quality and intelligibility metrics.
The backbone architecture is SoundStream. For WaveRVQ, the configuration is $(n_{\mathrm{iter}}, n_{\mathrm{code}}) = (3, 2)$, where the cumulative number of codebooks reaches $k=2$, 4, and 6 after the first, second, and third iterations, respectively.
The results show that all evaluation metrics improve as the number of codebooks increases for both RVQ and WaveRVQ.
When compared at the same number of codebooks, WaveRVQ achieves better SDR than conventional RVQ, indicating improved waveform reconstruction fidelity.
Moreover, WaveRVQ also improves perceptual quality metrics such as PESQ and UTMOS.
These results suggest that waveform-domain residual computation improves reconstruction accuracy without degrading perceptual quality.



\section{Downstream Task: SpeechLM-based TTS}
\label{section:speechlm}
To demonstrate the applicability of WaveRVQ to downstream tasks, we integrate it into SpeechLM-based TTS.
\subsection{Experimental Setting}
The system is implemented using ESPNet-SpeechLM~\cite{tian2025espnet}.
The multi-stream language model (LM) is based on a 24-layer Transformer decoder with 319M parameters and employs delay interleaving~\cite{copet2023simple}.
For the proposed method, we apply WaveRVQ to DAC with $(n_{\mathrm{iter}}, n_{\mathrm{code}}) = (3, 2)$. The RVQ-based DAC model is used as the baseline codec.
These codecs correspond to the models reported in Table~\ref{tab:other_architecture}.
The SpeechLM is trained on the 960-hour LibriSpeech~\cite{panayotov2015librispeech} dataset, and the test-clean subset is used for evaluation.
The generated speech is evaluated in terms of WER, speaker similarity, and UTMOS.
The speaker prompt is randomly selected from an utterance by the same speaker that is longer than 10 seconds, and the same prompt is used for all experiments.
The TTS system is trained for 200k steps on four NVIDIA H100 GPUs. During inference, we use top-$k$ sampling with $k=30$ and a sampling temperature of 1.2.

\subsection{Results}
As shown in Table~\ref{tab:speechlm}, the SpeechLM using the WaveRVQ-based codec outperforms the baseline model across all evaluation metrics. These results demonstrate the effectiveness of the proposed method for SpeechLM-based TTS.

Table~\ref{tab:speechlm_time} summarizes the real-time factor (RTF) of codec encoding, LM decoding, and codec decoding in the SpeechLM-based TTS system. The RTF is measured on an NVIDIA H100 GPU with a batch size of 1. As described in Section~\ref{section:wavervq}, the computational cost of WaveRVQ increases proportionally with the number of iterations.
This trend is reflected in Table~\ref{tab:speechlm_time}.
Regarding the encoding process, compared with the RVQ-based codec, the WaveRVQ-based codec requires an additional computational cost of two encoder-quantizer-decoder inference iterations because it is configured with $n_{\mathrm{iter}} = 3$.
In contrast, because the LM simultaneously generates codes for all iterations, the decoder can reconstruct the waveform components in parallel without depending on previous iterations.
As a result, its inference cost is comparable to that of the conventional RVQ-based codec.
Furthermore, LM decoding dominates the overall RTF, while codec processing accounts for only about 3.4\% of the total RTF. Moreover, the WaveRVQ-based model increases the total RTF by only 1.7\% compared with the RVQ-based model. Consequently, the additional computational overhead introduced by the proposed method is not a practical bottleneck for SpeechLM-based TTS.

\begin{table}[t]
    \centering
    \caption{SpeechLM-based TTS performance using DAC with RVQ and WaveRVQ on LibriSpeech test-clean subset.}
    \scalebox{0.8}{    
    \begin{tabular}{l|ccc}
    \toprule
        VQ Method & WER [\%] & SPK-Sim & UTMOS \\
        \midrule
        RVQ & 15.1 & 0.664 & 3.05 \\
        WaveRVQ & \textbf{10.8} & \textbf{0.739} & \textbf{3.28} \\        
        \bottomrule
    \end{tabular}}
    \label{tab:speechlm}
\end{table}

\begin{table}[t]
    \centering
    \caption{RTF breakdown for each component in the SpeechLM-based TTS on LibriSpeech test-clean subset.}
    \scalebox{0.8}{    
    \begin{tabular}{l|cccccc}
    \toprule
        VQ Method & Codec encoding & LLM & Codec decoding & Total \\            
        \midrule
        RVQ & 0.0038 & 0.568 & 0.0037 & 0.575  \\
        WaveRVQ & 0.0151 & 0.565 & 0.0045 & 0.585 \\        
        \bottomrule
    \end{tabular}}
    \label{tab:speechlm_time}
\end{table}

\section{Conclusion}
This paper presented WaveRVQ, which performs iterative residual computation in the waveform domain rather than in the latent space. The proposed framework reconstructs the target waveform as the sum of the waveform components generated at each iteration.
Experimental results demonstrated improvements in most evaluation metrics over conventional latent-space RVQ-based codecs for speech coding, joint speech coding and enhancement, and SpeechLM-based TTS.

\section*{Acknowledgment}
The authors used ChatGPT (OpenAI) solely for stylistic polishing and grammatical refinement of the manuscript.

\bibliographystyle{IEEEtran}
\bibliography{mybib}

@article{valin2012rfc,
  title={Definition of the Opus Audio Codec},
  author={Jean-Marc Valin and Koen Vos and Timothy B. Terriberry},
  journal={RFC},
  year={2012},
  volume={6716},
  pages={1-326},
}

@inproceedings{dietz2015overview,
  title={Overview of the {EVS} codec architecture},
  author={Dietz, Martin and Multrus, Markus and Eksler, Vaclav and Malenovsky, Vladimir and Norvell, Erik and Pobloth, Harald and Miao, Lei and Wang, Zhe and Laaksonen, Lasse and Vasilache, Adriana and others},
  booktitle={2015 IEEE International Conference on Acoustics, Speech and Signal Processing (ICASSP)},
  pages={5698--5702},
  year={2015},
  organization={IEEE}
}

@article{Morishima1990SpeechCB,
  title={Speech coding based on a multi-layer neural network},
  author={Shizuo Morishima and Hiroshi Harashima and Yasuo Katayama},
  journal={IEEE International Conference on Communications, Including Supercomm Technical Sessions},
  year={1990},
  pages={429-433 vol.2},
}

@inproceedings{kankanahalli2018end,
  title={End-To-End Optimized Speech Coding with Deep Neural Networks},
  author={Kankanahalli, Srihari},
  booktitle={2018 IEEE International Conference on Acoustics, Speech and Signal Processing (ICASSP)},
  pages={2521--2525},
  year={2018},
  organization={IEEE}
}

@article{van2017neural,
  title={Neural Discrete Representation Learning},
  author={Van Den Oord, Aaron and Vinyals, Oriol and others},
  journal={Advances in Neural Information Processing Systems},
  volume={30},
  year={2017}
}

@article{gray1984vector,
  title={Vector quantization},
  author={Gray, Robert},
  journal={IEEE Assp Magazine},
  volume={1},
  number={2},
  pages={4--29},
  year={1984},
  publisher={IEEE}
}

@article{vasuki2006review,
  title={A review of vector quantization techniques},
  author={Vasuki, A and Vanathi, Ponnusamy Thangapandian},
  journal={IEEE Potentials},
  volume={25},
  number={4},
  pages={39--47},
  year={2006},
  publisher={IEEE}
}

@article{
mousavi2025discrete,
title={Discrete Audio Tokens: More Than a Survey!},
author={Pooneh Mousavi and Gallil Maimon and Adel Moumen and Darius Petermann and Jiatong Shi and Haibin Wu and Haici Yang and Anastasia Kuznetsova and Artem Ploujnikov and Ricard Marxer and Bhuvana Ramabhadran and Benjamin Elizalde and Loren Lugosch and Jinyu Li and Cem Subakan and Phil Woodland and Minje Kim and Hung-yi Lee and Shinji Watanabe and Yossi Adi and Mirco Ravanelli},
journal={Transactions on Machine Learning Research},
issn={2835-8856},
year={2025},
note={}
}

@inproceedings{wu2024codec,
  title={{Codec-SUPERB}: An In-Depth Analysis of Sound Codec Models},
  author={Wu, Haibin and Chung, Ho-Lam and Lin, Yi-Cheng and Wu, Yuan-Kuei and Chen, Xuanjun and Pai, Yu-Chi and Wang, Hsiu-Hsuan and Chang, Kai-Wei and Liu, Alex and Lee, Hung-yi},
  booktitle={Findings of the Association for Computational Linguistics: ACL 2024},
  pages={10330--10348},
  year={2024}
}

@article{
arora2025on,
title={On The Landscape of Spoken Language Models: A Comprehensive Survey},
author={Siddhant Arora and Kai-Wei Chang and Chung-Ming Chien and Yifan Peng and Haibin Wu and Yossi Adi and Emmanuel Dupoux and Hung-yi Lee and Karen Livescu and Shinji Watanabe},
journal={Transactions on Machine Learning Research},
issn={2835-8856},
year={2025},
}

@inproceedings{cui-etal-2025-recent,
    title = {Recent Advances in Speech Language Models: A Survey},
    author = "Cui, Wenqian  and
      Yu, Dianzhi  and
      Jiao, Xiaoqi  and
      Meng, Ziqiao  and
      Zhang, Guangyan  and
      Wang, Qichao  and
      Guo, Steven Y.  and
      King, Irwin",    
    booktitle = "Proceedings of the 63rd Annual Meeting of the Association for Computational Linguistics (Volume 1: Long Papers)",
    month = jul,
    year = "2025",    
    publisher = "Association for Computational Linguistics",    
    pages = "13943--13970",
}

@article{zeghidour2021soundstream,
  title={{SoundStream}: An End-to-End Neural Audio Codec},
  author={Zeghidour, Neil and Luebs, Alejandro and Omran, Ahmed and Skoglund, Jan and Tagliasacchi, Marco},
  journal={IEEE/ACM Transactions on Audio, Speech, and Language Processing},
  volume={30},
  pages={495--507},
  year={2021},
  publisher={IEEE}
}

@article{defossez2022high,
title={High Fidelity Neural Audio Compression},
author={Alexandre D{\'e}fossez and Jade Copet and Gabriel Synnaeve and Yossi Adi},
journal={Transactions on Machine Learning Research},
issn={2835-8856},
year={2023},
}

@article{kumar2023high,
  title={High-Fidelity Audio Compression with Improved {RVQGAN}},
  author={Kumar, Rithesh and Seetharaman, Prem and Luebs, Alejandro and Kumar, Ishaan and Kumar, Kundan},
  journal={Advances in Neural Information Processing Systems},
  volume={36},
  pages={27980--27993},
  year={2023}
}

@inproceedings{zhang2024speechtokenizer,
  title={Speechtokenizer: Unified speech tokenizer for speech language models},
  author={Zhang, Xin and Zhang, Dong and Li, Shimin and Zhou, Yaqian and Qiu, Xipeng},
  booktitle={International Conference on Learning Representations},
  volume={2024},
  pages={31798--31818},
  year={2024}
}

@inproceedings{ji2025wavtokenizer,
  title={Wavtokenizer: an efficient acoustic discrete codec tokenizer for audio language modeling},
  author={Ji, Shengpeng and Jiang, Ziyue and Wang, Wen and Chen, Yifu and Fang, Minghui and Zuo, Jialong and Yang, Qian and Cheng, Xize and Li, Ruiqi and Zhang, Ziang and others},
  booktitle={International Conference on Learning Representations},
  volume={2025},
  pages={93809--93826},
  year={2025}
}

@InProceedings{pmlr-v235-ju24b,
  title = 	 {{N}atural{S}peech 3: Zero-Shot Speech Synthesis with Factorized Codec and Diffusion Models},
  author =       {Ju, Zeqian and Wang, Yuancheng and Shen, Kai and Tan, Xu and Xin, Detai and Yang, Dongchao and Liu, Eric and Leng, Yichong and Song, Kaitao and Tang, Siliang and Wu, Zhizheng and Qin, Tao and Li, Xiangyang and Ye, Wei and Zhang, Shikun and Bian, Jiang and He, Lei and Li, Jinyu and Zhao, Sheng},
  booktitle = 	 {Proceedings of the 41st International Conference on Machine Learning},
  pages = 	 {22605--22623},
  year = 	 {2024},  
  volume = 	 {235},
  series = 	 {Proceedings of Machine Learning Research},
  month = 	 {21--27 Jul},
  publisher =    {PMLR},
}

@article{bengio2013estimating,
  title={Estimating or Propagating Gradients Through Stochastic Neurons for Conditional Computation},
  author={Bengio, Yoshua and L{\'e}onard, Nicholas and Courville, Aaron},
  journal={arXiv preprint arXiv:1308.3432},
  year={2013}
}

@article{borsos2023audiolm,
  title={{AudioLM}: A Language Modeling Approach to Audio Generation},
  author={Zal{\'a}n Borsos and Rapha{\"e}l Marinier and Damien Vincent and Eugene Kharitonov and Olivier Pietquin and Matthew Sharifi and Dominik Roblek and Olivier Teboul and David Grangier and Marco Tagliasacchi and Neil Zeghidour},
  journal={IEEE/ACM Transactions on Audio, Speech, and Language Processing},
  year={2022},
  volume={31},
  pages={2523-2533},  
}

@inproceedings{tian2025espnet,
  title={{ESPnet-SpeechLM}: An Open Speech Language Model Toolkit},
  author={Tian, Jinchuan and Shi, Jiatong and Chen, William and Arora, Siddhant and Masuyama, Yoshiki and Maekaku, Takashi and Wu, Yihan and Peng, Junyi and Bharadwaj, Shikhar and Zhao, Yiwen and others},
  booktitle={Proceedings of the 2025 Conference of the Nations of the Americas Chapter of the Association for Computational Linguistics: Human Language Technologies (System Demonstrations)},
  pages={116--124},
  year={2025}
}

@article{wang2023neural,
  title={Neural Codec Language Models are Zero-Shot Text to Speech Synthesizers},
  author={Chengyi Wang and Sanyuan Chen and Yu Wu and Zi-Hua Zhang and Long Zhou and Shujie Liu and Zhuo Chen and Yanqing Liu and Huaming Wang and Jinyu Li and Lei He and Sheng Zhao and Furu Wei},
  journal={IEEE Transactions on Audio, Speech and Language Processing},
  year={2023},
  volume={33},
  pages={705-718},
}

@article{chen2024vall,
  title={Vall-e 2: Neural codec language models are human parity zero-shot text to speech synthesizers},
  author={Chen, Sanyuan and Liu, Shujie and Zhou, Long and Liu, Yanqing and Tan, Xu and Li, Jinyu and Zhao, Sheng and Qian, Yao and Wei, Furu},
  journal={arXiv preprint arXiv:2406.05370},
  year={2024}
}

@article{du2024cosyvoice,
  title={{CosyVoice 2}: Scalable Streaming Speech Synthesis with Large Language Models},
  author={Du, Zhihao and Wang, Yuxuan and Chen, Qian and Shi, Xian and Lv, Xiang and Zhao, Tianyu and Gao, Zhifu and Yang, Yexin and Gao, Changfeng and Wang, Hui and others},
  journal={arXiv preprint arXiv:2412.10117},
  year={2024}
}

@inproceedings{gu2024esc,
  title={{ESC}: Efficient Speech Coding with Cross-Scale Residual Vector Quantized Transformers},
  author={Gu, Yuzhe and Diao, Enmao},
  booktitle={Proceedings of the 2024 Conference on Empirical Methods in Natural Language Processing},
  pages={10084--10096},
  year={2024}
}

@article{yang2023hifi,
  title={{HiFi-Codec}: Group-residual Vector quantization for High Fidelity Audio Codec},
  author={Yang, Dongchao and Liu, Songxiang and Huang, Rongjie and Tian, Jinchuan and Weng, Chao and Zou, Yuexian},
  journal={arXiv preprint arXiv:2305.02765},
  year={2023}
}

@inproceedings{
siuzdak2024snac,
title={{SNAC}: Multi-Scale Neural Audio Codec},
author={Hubert Siuzdak and Florian Gr{\"o}tschla and Luca A Lanzend{\"o}rfer},
booktitle={Audio Imagination: NeurIPS 2024 Workshop AI-Driven Speech, Music, and Sound Generation},
year={2024},
}

@inproceedings{zhen2019cascaded,
  author       = {Kai Zhen and
                  Jongmo Sung and
                  Mi Suk Lee and
                  Seungkwon Beack and
                  Minje Kim},  
  title        = {Cascaded Cross-Module Residual Learning Towards Lightweight End-to-End
                  Speech Coding},
  booktitle    = {20th Annual Conference of the International Speech Communication Association,
                  Interspeech 2019, Graz, Austria, September 15-19, 2019},
  pages        = {3396--3400},
  publisher    = {{ISCA}},
  year         = {2019},  
}

@inproceedings{zhen2020efficient,
  title={Efficient and Scalable Neural Residual Waveform Coding with Collaborative Quantization},
  author={Zhen, Kai and Lee, Mi Suk and Sung, Jongmo and Beack, Seungkwon and Kim, Minje},
  booktitle={ICASSP 2020-2020 IEEE International Conference on Acoustics, Speech and Signal Processing (ICASSP)},
  pages={361--365},
  year={2020},
  organization={IEEE}
}

@article{zhen2021scalable,
  title={Scalable and efficient neural speech coding: A hybrid design},
  author={Zhen, Kai and Sung, Jongmo and Lee, Mi Suk and Beack, Seungkwon and Kim, Minje},
  journal={IEEE/ACM Transactions on Audio, Speech, and Language Processing},
  volume={30},
  pages={12--25},
  year={2021},
  publisher={IEEE}
}

@article{wojcicki2025low,
  title={Low-Resource Audio Codec {(LRAC)}: 2025 Challenge Description},
  author={Wojcicki, Kamil and Isik, Yusuf Ziya and Lechler, Laura and Yesilbursa, Mansur and Bali{\'c}, Ivana and Mack, Wolfgang and {\L}aganowski, Rafa{\l} and Zhang, Guoqing and Adi, Yossi and Kim, Minje and others},
  journal={arXiv preprint arXiv:2510.23312},
  year={2025}
}

@article{isik2025baseline,
  title={Baseline Systems For The 2025 Low-Resource Audio Codec Challenge},
  author={Isik, Yusuf Ziya and {\L}aganowski, Rafa{\l}},
  journal={arXiv preprint arXiv:2510.00264},
  year={2025}
}

@inproceedings{yamamoto2020parallel,
  title={{Parallel Wavegan}: A Fast Waveform Generation Model Based on Generative Adversarial Networks with Multi-Resolution Spectrogram},
  author={Yamamoto, Ryuichi and Song, Eunwoo and Kim, Jae-Min},
  booktitle={ICASSP 2020-2020 IEEE International Conference on Acoustics, Speech and Signal Processing (ICASSP)},
  pages={6199--6203},
  year={2020},
  organization={IEEE}
}

@inproceedings{mao2017least,
  title={Least Squares Generative Adversarial Networks},
  author={Mao, Xudong and Li, Qing and Xie, Haoran and Lau, Raymond YK and Wang, Zhen and Paul Smolley, Stephen},
  booktitle={Proceedings of the IEEE International Conference on Computer Vision},
  pages={2794--2802},
  year={2017}
}

@inproceedings{shi2025versa,
  title={{VERSA}: A versatile evaluation toolkit for speech, audio, and music},
  author={Shi, Jiatong and Shim, Hye-jin and Tian, Jinchuan and Arora, Siddhant and Wu, Haibin and Petermann, Darius and Yip, Jia Qi and Zhang, You and Tang, Yuxun and Zhang, Wangyou and others},
  booktitle={Proceedings of the 2025 Conference of the Nations of the Americas Chapter of the Association for Computational Linguistics: Human Language Technologies (System Demonstrations)},
  pages={191--209},
  year={2025}
}

@inproceedings{rix2001perceptual,
  title={Perceptual evaluation of speech quality {(PESQ)}-a new method for speech quality assessment of telephone networks and codecs},
  author={Rix, Antony W and Beerends, John G and Hollier, Michael P and Hekstra, Andries P},
  booktitle={2001 IEEE International Conference on Acoustics, Speech, and Signal Processing. Proceedings (Cat. No. 01CH37221)},
  volume={2},
  pages={749--752},
  year={2001},
  organization={IEEE}
}

@inproceedings{saeki2022utmos,
  title={{UTMOS: UTokyo-SaruLab System for VoiceMOS Challenge 2022}},
  author={Saeki, Takaaki and Xin, Detai and Nakata, Wataru and Koriyama, Tomoki and Takamichi, Shinnosuke and Saruwatari, Hiroshi},
  booktitle={Proc. Interspeech 2022},
  pages={4521--4525},
  month = {09},
  year={2022}
}

@article{ragano2024scoreq,
  title={{SCOREQ}: Speech quality assessment with contrastive regression},
  author={Ragano, Alessandro and Skoglund, Jan and Hines, Andrew},
  journal={Advances in Neural Information Processing Systems},
  volume={37},
  pages={105702--105729},
  year={2024}
}

@article{huang2024mos,
  title={{MOS-Bench}: Benchmarking generalization abilities of subjective speech quality assessment models},
  author={Huang, Wen-Chin and Cooper, Erica and Toda, Tomoki},
  journal={arXiv preprint arXiv:2411.03715},
  year={2024}
}

@article{tjandra2025meta,
  title={Meta audiobox aesthetics: Unified automatic quality assessment for speech, music, and sound},
  author={Tjandra, Andros and Wu, Yi-Chiao and Guo, Baishan and Hoffman, John and Ellis, Brian and Vyas, Apoorv and Shi, Bowen and Chen, Sanyuan and Le, Matt and Zacharov, Nick and others},
  journal={arXiv preprint arXiv:2502.05139},
  year={2025}
}

@inproceedings{radford2023robust,
  title={Robust speech recognition via large-scale weak supervision},
  author={Radford, Alec and Kim, Jong Wook and Xu, Tao and Brockman, Greg and McLeavey, Christine and Sutskever, Ilya},
  booktitle={International conference on machine learning},
  pages={28492--28518},
  year={2023},
  organization={PMLR}
}

@techreport{ITU-T_P800_1996,
  author      = "{ITU-T}",
  title       = "{Methods for Subjective Determination of Transmission Quality}",  
  type        = "Recommendation",
  number      = "P.800",
  month       = aug,
  year        = "1996"
}

@inproceedings{watanabe2018espnet,
  author={Shinji Watanabe and Takaaki Hori and Shigeki Karita and Tomoki Hayashi and Jiro Nishitoba and Yuya Unno and Nelson {Enrique Yalta Soplin} and Jahn Heymann and Matthew Wiesner and Nanxin Chen and Adithya Renduchintala and Tsubasa Ochiai},
  title={{ESPnet}: End-to-End Speech Processing Toolkit},
  year={2018},
  booktitle={Proceedings of Interspeech},
  pages={2207--2211},
  doi={10.21437/Interspeech.2018-1456},
  url={http://dx.doi.org/10.21437/Interspeech.2018-1456}
}

@inproceedings{shi2024espnet,
  title={{ESPnet-Codec}: Comprehensive training and evaluation of neural codecs for audio, music, and speech},
  author={Shi, Jiatong and Tian, Jinchuan and Wu, Yihan and Jung, Jee-weon and Yip, Jia Qi and Masuyama, Yoshiki and Chen, William and Wu, Yuning and Tang, Yuxun and Baali, Massa and others},
  booktitle={2024 IEEE Spoken language technology workshop (SLT)},
  pages={562--569},
  year={2024},
  organization={IEEE}
}

@article{jung2024espnet,
  title={{ESPnet-SPK}: full pipeline speaker embedding toolkit with reproducible recipes, self-supervised front-ends, and off-the-shelf models},
  author={Jung, Jee-weon and Zhang, Wangyou and Shi, Jiatong and Aldeneh, Zakaria and Higuchi, Takuya and Theobald, Barry-John and Abdelaziz, Ahmed Hussen and Watanabe, Shinji},
  journal={Proc. Interspeech 2024},
  pages={4278--4282},
  year={2024}
}

@article{jung2022pushing,
  title={Pushing the limits of raw waveform speaker recognition},
  author={Jung, Jee-weon and Kim, You Jin and Heo, Hee-Soo and Lee, Bong-Jin and Kwon, Youngki and Chung, Joon Son},
  journal={Proc. Interspeech 2022},
  pages={2228--2232},
  year={2022}
}

@article{zwicker1961subdivision,
  title={Subdivision of the audible frequency range into critical bands (Frequenzgruppen)},
  author={Zwicker, Eberhard},
  journal={The Journal of the Acoustical Society of America},
  volume={33},
  number={2},
  pages={248--248},
  year={1961},
  publisher={Acoustical Society of America}
}

@article{glasberg1990derivation,
  title={Derivation of auditory filter shapes from notched-noise data},
  author={Glasberg, Brian R and Moore, Brian CJ},
  journal={Hearing Research},
  volume={47},
  number={1-2},
  pages={103--138},
  year={1990},
  publisher={Elsevier}
}

@article{copet2023simple,
  title={Simple and controllable music generation},
  author={Copet, Jade and Kreuk, Felix and Gat, Itai and Remez, Tal and Kant, David and Synnaeve, Gabriel and Adi, Yossi and D{\'e}fossez, Alexandre},
  journal={Advances in neural information processing systems},
  volume={36},
  pages={47704--47720},
  year={2023}
}

@inproceedings{panayotov2015librispeech,
  title={Librispeech: an asr corpus based on public domain audio books},
  author={Panayotov, Vassil and Chen, Guoguo and Povey, Daniel and Khudanpur, Sanjeev},
  booktitle={2015 IEEE international conference on acoustics, speech and signal processing (ICASSP)},
  pages={5206--5210},
  year={2015},
  organization={IEEE}
}

@article{leung2022speaking,
  title={Speaking fundamental frequencies of adult speakers of Australian English and effects of sex, age, and geographical location},
  author={Leung, Yeptain and Oates, Jennifer and Papp, Vikt{\'o}ria and Chan, Siew-Pang},
  journal={Journal of Voice},
  volume={36},
  number={3},
  pages={434--e1},
  year={2022},
  publisher={Elsevier}
}

\end{document}